\documentclass[twocolumn]{IEEEtran} 

\usepackage[pdftex]{graphicx}

\DeclareGraphicsExtensions{.png,.tif}

\begin{document}

\title{Expansion of a plasma cloud into the solar wind} 
\author{ L. Gargat\'e, R. A. Fonseca, R. Bingham, L. O. Silva
 \thanks{L. Gargat\'{e}, R. A. Fonseca and L. O. Silva are with GoLP/Instituto de Plasmas e Fus\~ao Nuclear, Instituto 
Superior T\'ecnico, 1049-001 Lisboa, Portugal 
(e-mail: luisgargate@ist.utl.pt)}
\thanks{R. Bingham is with SSTD, Rutherford Appleton Laboratory, Didcot, Oxon, OX11 0QX UK}
 }

\maketitle

\begin{abstract}
Three-dimensional (3D) hybrid particle-in-cell (PIC) simulations, with kinetic ions and fluid electrons, of a plasma cloud expansion in the solar wind are presented, revealing the dynamics of the expansion, with shock formation, magnetic field compression, and the solar wind ions deflection around the plasma bubble. The similarities of this system with a magnetosphere are also pointed out.
\end{abstract}


%
%
The use of artificial magnetospheres is now under study as a possible protection mechanism for spacecrafts in the interplanetary space environment. Experiments in the laboratory are now being performed \cite{experiments} to study the interaction and the deflection of an incoming plasma beam by: i) a standing dipolar magnetic field, ii) a plasma cloud expanding into the plasma beam, and iii) a dipolar magnetic field and a plasma source forming an expanding plasma. The plasma density in the laboratory is $10^{18}\,\mathrm{m^{-3}}$ with a temperature of $5\,\mathrm{eV}$ and a velocity of $400\,\mathrm{km/s}$, yielding a low plasma $\beta=0.05$, an acoustic Mach number of $M_{cs}\sim13$ and an Alfv\`{e}nic Mach number of $M_{ca}\sim0.9$. For the solar wind present in interplanetary space, the velocity is in the order of $400\,\mathrm{km/s}$, the density is $\sim 5\,\mathrm{cm^{-3}}$, and the temperature is $\sim 20\,\mathrm{eV}$, yielding $\beta\sim 1$, $M_{cs}\sim7$ and $M_{ca}\sim 4$. This leads to substantial physical differences between the two cases that can be bridged via computer simulations.

Experiments in a space plasma environment \cite{ampte}, i.e. the expansion of a plasma cloud into the solar wind ( scenario ii)), have shown several distinct characteristics. In this paper, we illustrate the 3D features of this system. Numerical simulations are performed using \textit{dHybrid} \cite{gargate}, a hybrid code with kinetic ions and fluid electrons. In the hybrid approximation, the displacement current in Amp\`{e}re's law is neglected, quasi-neutrality is assumed, and moments of the Vlasov equation for the electrons are calculated in order to obtain the generalized Ohm's law, that yields the electric field \cite{lipatov}. In our simulations, the solar wind propagates in the $+x$ direction, magnetic field is oriented in the $+z$ direction and the cloud expands in the middle of the simulation box. The box size is $\left(\sim 15000\,\mathrm{km}\right)^3$ corresponding to $\left(\sim 41\,\mathrm{r_{Li}}\right)^3$ (ion Larmor radius) of the solar wind. The time step is $0.01\,\mathrm{s}$ corresponding to $3.5\times10^{-4}\,\mathrm{T_{ci}}$ (ion gyro periods) and $\left(300\right)^3$ cells are used, with $8$ particles per cell, yielding a spatial resolution of $0.14\,\mathrm{r_{Li}}$. 

\begin{figure}[htb]
\centering
\includegraphics[width=3.6in]{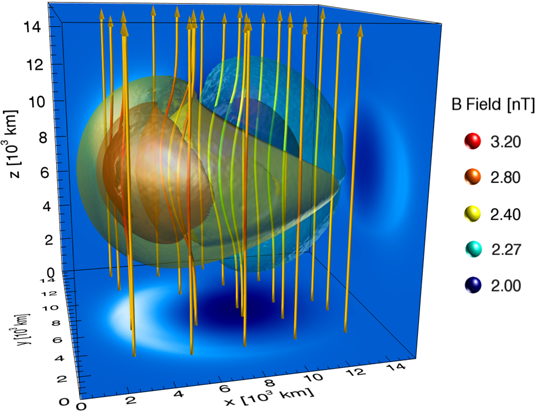}
\caption{Magnetic field iso-surfaces, field lines and projections for $t=80\,\mathrm{s}$ corresponding to $2.81\,\mathrm{T_{ci}}$. Box size is $\sim \left(150\,\mathrm{c/\omega_{pi}}\right)^3$.}
\label{figure1}
\end{figure}

\begin{figure}[htb]
\centering
\includegraphics[width=3.6in]{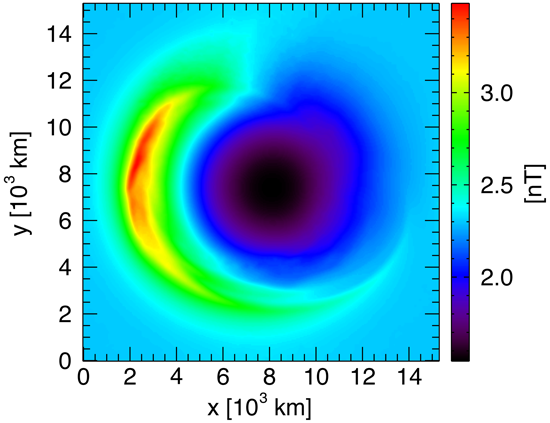}
\caption{Slice of the magnetic field intensity in the $xy$ plane, in the middle of the simulation box.}
\label{figure2}
\end{figure}

A shock front between the incoming solar wind and the cloud is formed; the magnetic field lines are pushed out of the cloud region by the expanding plasma cloud, Fig. \ref{figure1}. The shock front is characterized by a compression of the magnetic field in the shock region, as can be seen in both Fig. \ref{figure1} and Fig. \ref{figure2}. The magnetic barrier created is larger than the volume occupied by the bulk of the plasma cloud; a similar scenario has also been observed when a dipole magnetic field, and hence a magnetosphere, is present. Furthermore, the expanding plasma cloud creates a magnetic field void in the cloud zone, i.e. a diamagnetic cavity, Fig. \ref{figure2}.

\begin{figure}[htb]
\centering
\includegraphics[width=3.6in]{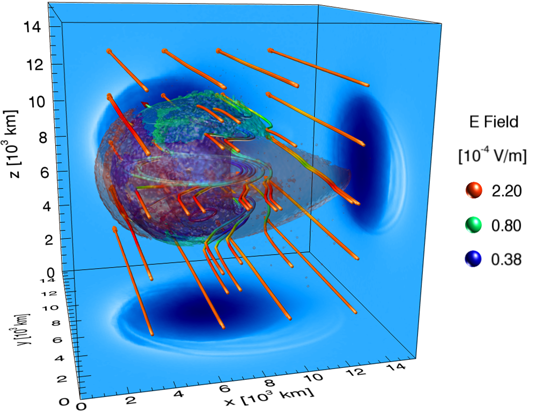}
\caption{Electric field iso-surfaces, field lines and projections for $t=80\,\mathrm{s}$. Solar wind is coming from the left in the $+x$ direction. Magnetic field is in the $+z$ direction.}
\label{figure3}
\end{figure}

\begin{figure}[htb]
\centering
\includegraphics[width=3.6in]{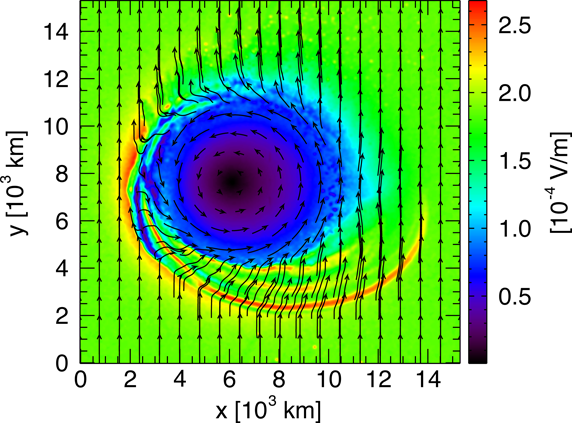}
\caption{Slice of the electric field in the $xy$ plane, in the middle of the simulation box, showing the electric field intensity and field lines.}
\label{figure4}
\end{figure}

The electric field is quasi-perpendicular to the magnetic field, the field lines taking a counter clockwise orientation due to the velocity of the outward expanding plasma ions, cf. Fig. \ref{figure3} and Fig. \ref{figure4}. The cloud ions expand due to their thermal velocity. The faster ions are, after some time, further away from the cloud center producing an electric field, seen in Fig. \ref{figure4}, that increases in intensity with increasing radius due to the increase in velocity with increasing radius. Such field is responsible also for the ejection of cloud ions in the $-\vec{v}\times\vec{B}$ side of the cloud and the preferred solar wind deflection direction: the solar wind is deflected around the magnetic/plasma barrier, flowing preferentially around the cloud in the $\vec{v}\times\vec{B}$ side of the cloud.
 
In conclusion, three-dimensional hybrid simulations of a space plasma experiment can now capture the relevant physics associated with the expansion of a plasma cloud in the solar wind flow, demonstrating the magnetic field compression (due to the shock), the magnetic field expansion and the increased interaction region due to the magnetic field. These features are similar to the typical case where a magnetosphere is present. Our results illustrate the key physics ruling this scenario, and show the possibility of deflecting the solar wind using the dynamics of a plasma cloud, thus providing benchmarks for the tools required for the interpretation of the ongoing laboratory experiments.
 
\section*{Acknowledgments}
This work was supported by grants SFRH/BD/17750/2004 and POCI/66823/2006 from Funda\c{c}\~{a}o para a Ci\^{e}ncia e a Tecnologia (Portugal). The simulations presented here were produced using the IST Cluster (IST/Portugal).


\nocite{*}
\bibliographystyle{IEEE}

%
\end{document}